\documentclass[12pt,preprint]{aastex}
\slugcomment{to be published in Astrophysical Journal Letters}
\lefthead{Liang et al.} \righthead{ATOMIC HYDROGEN IN HD~209458B}
\begin{document}
\title{Source of Atomic Hydrogen in the Atmosphere of HD~209458b}
\author{Mao-Chang Liang, Christopher D. Parkinson, Anthony Y.-T. Lee, Yuk L. Yung}
\affil{Division of Geological and Planetary Sciences, California
Institute of Technology, 1201 E. California Blvd., Pasadena, CA
91125} \email{mcl@gps.caltech.edu, cdp@gps.caltech.edu,
ytl@gps.caltech.edu, yly@gps.caltech.edu} \and
\author{Sara Seager}
\affil{Department of Terrestrial Magnetism, Carnegie Institution
of Washington, 5241 Broad Branch Rd. NW Washington, D.C. 20015}
\email{seager@dtm.ciw.edu}

\begin{abstract}
Atomic hydrogen loss at the top of HD~209458b's atmosphere has
been recently detected \citep{VM03}. We have developed a
1-dimensional model to study the chemistry in the upper atmosphere
of this extrasolar ``hot jupiter". The 3 most abundant elements
(other than He), as well as 4 parent molecules are included in
this model, viz., H, C, O, H$_2$, CO, H$_2$O, and CH$_4$. The
higher temperatures ($\sim 1000$~K) and higher stellar irradiance
($\sim 6 \times 10^5$~W~m$^{-2}$) strongly enhance and modify the
chemical reaction rates in this atmosphere. Our two main results
are that (a) the production of atomic hydrogen in the atmosphere
is mainly driven by H$_2$O photolysis and reaction of OH with
H$_2$, and is not sensitive to the exact abundances of CO, H$_2$O,
and CH$_4$, and (b) H$_2$O and CH$_4$ can be produced via the
photolysis of CO followed by the reactions with H$_2$.

\end{abstract}


\section{INTRODUCTION}

Since the discovery of the first extrasolar planet, 51~Peg~b, in
1995 \citep{MQ95}, a total of 102 planets have so far been
discovered (e.g., Butler et al. 2003; Udry et al. 2002, and
references therein) and analyzed statistically in order to
characterize the formation environment \citep{Fetal02,SIMRU03}.
The formation of gas giants is thought to be complete in 10~Myr,
before the disappearance of the gaseous stellar accretion disk, at
distances $> 5$~AU from the parent star. They are then pulled to
their present positions by tidal interaction between the gas disk
and planet (e.g., Pollack et al. 1996; Ward 1997).

An edge-on planet provides a unique opportunity to investigate the
planetary atmosphere. HD~209458b is such a planet, providing the
first extrasolar planetary detection using the light curve
obtained during a planetary transit of its parent star
\citep{CBLM00,HMBV00}. The orbital parameters were accurately
determined by \citet{CBLM00}, \citet{HMBV00}, and \citet{Metal00}.
Strong absorption lines are required to make an atmospheric
detection and \citet{SS00} theoretically characterized the most
prominent absorption features, viz., \ion{Na}{1} and \ion{K}{1}
doublet resonance and \ion{He}{1}~$2^3S-2^3P$ triplets.
\citet{CGNG02} detected the \ion{Na}{1} doublet at 589.3~nm in
HD~209458b with $\sim 4\sigma$ confidence level. Following this,
\citet{VM03} made the first observation of the extended upper
atmosphere of HD~209458b with a $\sim 4 \sigma$ detection of the
\ion{H}{1} atomic hydrogen absorption of the stellar
Lyman-$\alpha$ line. They reported an absorption of $\sim 15 \pm
4$\%  and claimed this should be taking place beyond the Roche
limit, thus implying hydrodynamic escape of hydrogen atoms from
HD~209458b's atmosphere.

The temperature and UV flux of close-in planets are high. This
motivates us to study the chemistry that may be important in this
"hot jupiter". In this paper, we consider a simple
hydrocarbon/oxygen chemistry model to determine a source of atomic
hydrogen in the atmosphere of HD~209458b and represents the first
effort to investigate the UV enhanced chemical processes in "hot
jupiters".

\section{MODEL}
Our model is based on the four parent molecules H$_2$, CO, H$_2$O,
and CH$_4$ and is a derivative of the Caltech/JPL KINETICS model
for the Jovian atmosphere. HD~209458b is orbiting at a distance of
0.05 AU. As HD~209458 is a G0 solar-type dwarf star, it is
justified to use the solar spectrum. We expect the atmosphere to
have a temperature, $\sim 1000$ K, and UV flux ($< 1800$~\AA),
$\sim 2\times 10^{15}$~photons~cm$^{-2}$~s$^{-1}$. By comparison
the UV flux at Jupiter is a factor of $\sim 10^4$ lower.

{\it Hydrocarbons} The hydrocarbon photochemical scheme used here
is a simplified version of the Jovian atmospheric model described
in \citet{S73} and \citet{Getal96}. The photodissociation of
CH$_4$ and the subsequent reactions of the species with hydrogen
produce all the other hydrocarbons present in a Jovian-type
atmosphere. For lower temperatures and weaker stellar irradiation,
the main source of H is from H$_2$ and CH$_4$ photodissociation
and the main sink via C$_2$H$_2$, which acts as a catalyst in
recombining H. HD~209458b receives much greater stellar
irradiation and is therefore much hotter than Jupiter. In this
case, the formation of H is greatly enhanced by photolysis of
H$_2$O and reactions between O and OH radicals and H$_2$. The sink
for H is more complex (see $\S$~\ref{H_H2O}).

{\it Oxygen} O is similar in abundance to C and represents a
cosmic abundance of these species. The atmospheric H$_2$O
abundance will be controlled mainly by the comparative richness of
these two species. The amount of H$_2$O and CO governs the amount
of atomic oxygen present and the related reactions are of great
interest. The oxygen related reactions are taken from Moses et al.
(2000).

\begin{figure*}
\epsscale{1} \plotone{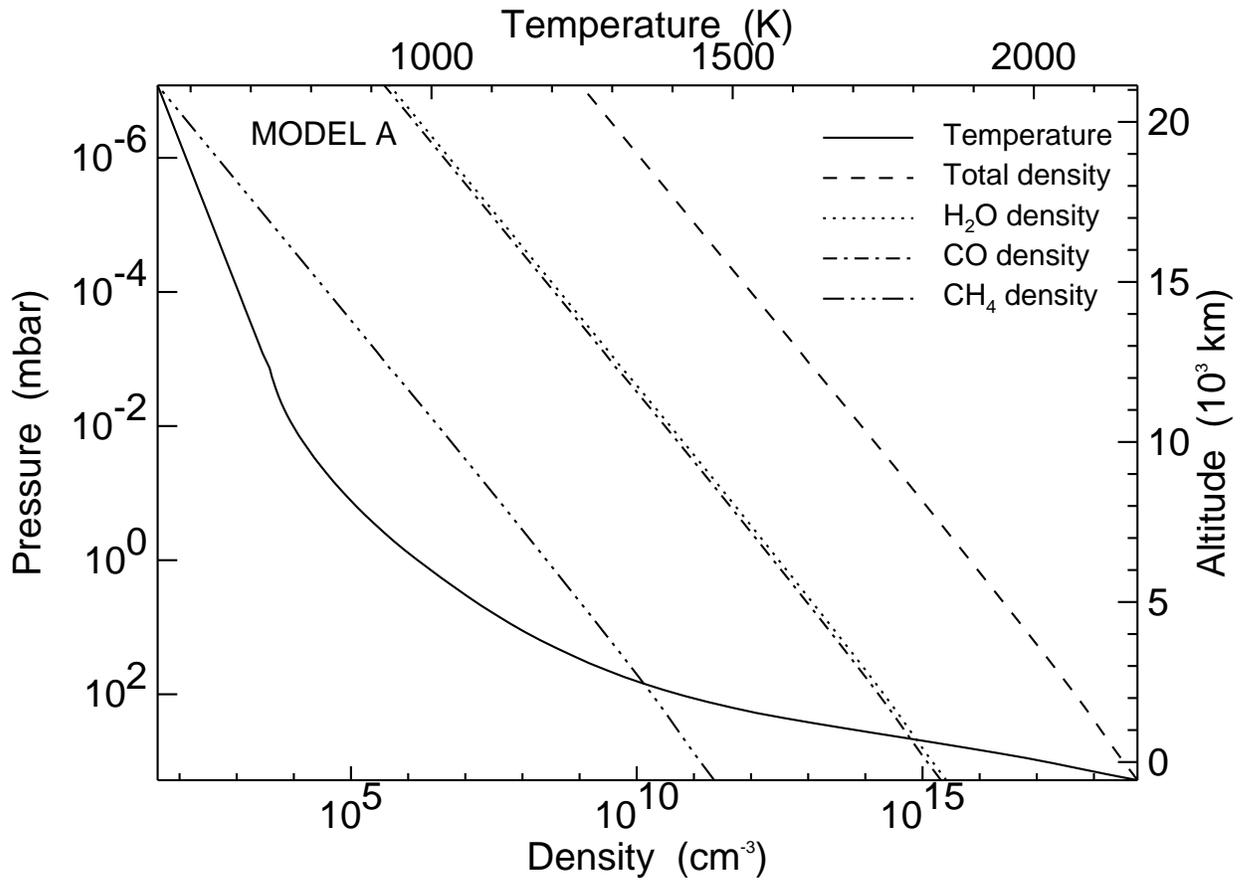} \caption[profiles of
atmosphere]{Vertical profile of temperature, total density, and
constituent number density. \label{profile}}
\end{figure*}

{\it Model atmosphere} Three models are investigated in this
paper. Our standard reference model using solar abundances is
shown in Fig.~\ref{profile} (Model A). We also consider 2 other
cases, Model B and Model C, in which H$_2$O and CO abundances are
10 times lower, respectively. We have taken the 1 bar level to be
"0" km, and all heights are referenced from this level. The
temperature-pressure profile and chemical abundances are based on
\citet{Setal00}. The temperature decreases from the bottom to the
top of the atmosphere. The abundances of CO and H$_2$O are,
$3.6\times 10^{-4}$ and $4.5\times 10^{-4}$, respectively. These
values are similar to solar abundances. The CH$_4$ abundance is
determined by the thermodynamic equilibrium chemistry in the deep
atmosphere. We adopt the value $3.9\times 10^{-8}$, which is the
lower end of the model by \citet{SS00}. The temperature-pressure
profiles are not certain, because global circulation and high
temperature condensation \citep{SS00,SBP00} are not included in
generating the model atmosphere. Nevertheless, the present
standard reference model is accurate enough for a first order
understanding of the chemistry and characterizes a source of H in
the upper atmosphere. The eddy diffusion is proportional to
n$^{-\alpha}$ (n is number density) where $\alpha$ is taken to be
$\sim 0.6-0.7$.

\section{RESULTS}
The principle results for our three models will be described in
three sections. Section \ref{OH_O} deals with OH and O radicals,
followed by CO$_2$ and CH$_4$ in section \ref{CO2_CH4}. The
important question of hydrogen production in relation to H$_2$O is
addressed in section \ref{H_H2O}.

\subsection{OH and O radicals \label{OH_O}}
\begin{figure*}
\epsscale{0.7} \plotone{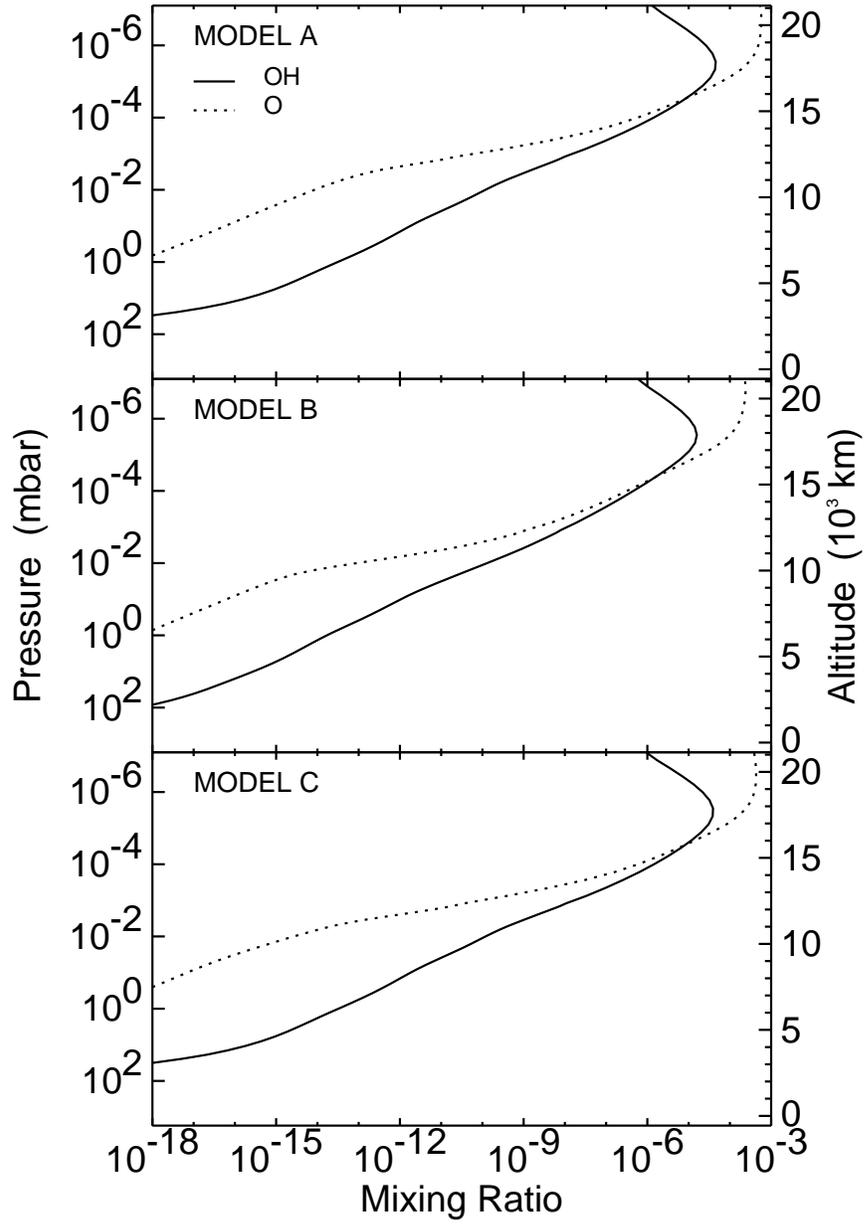} \caption[resulting mixing ratios of
radicals]{Comparison of mixing ratios of OH and O radicals in
Models A, B, and C. \label{radicals}}
\end{figure*}

Fig.~\ref{radicals} shows the OH and O radicals in our models. OH
and O are the most important radicals as they drive most of the
chemical reactions. O is the most important element in
facilitating the formation of OH radicals in the water-poor
atmosphere. These species are produced photochemically, for
example,
$$ \mbox{CO} + \mbox{h$\nu$} \rightarrow \mbox{C} + \mbox{O}, $$
$$ \mbox{H$_2$O} + \mbox{h$\nu$} \rightarrow \mbox{H} + \mbox{OH}. $$
CO photolysis is an important source of O and H$_2$O is the main
source of OH. Though O($^1$D) is not as abundant as O ($\lesssim
10^{-5}$~[O]), it can react with H$_2$ to produce a similar amount
of OH. O is increasing with altitude as a consequence of H$_2$O
and CO photolysis. OH is increasing with altitude until it starts
decreasing at $\sim 10$~nbar. The decline of OH above 10~nbar is
due to OH photodissociation. We see the mixing ratio of OH
radicals is not sensitive to the abundance of CO and H$_2$O. With
an order of magnitude change in H$_2$O (Model B) or CO (Model C),
OH is changed only by a factor of $\lesssim 3$. However, O is
sensitive to both CO and H$_2$O concentrations and preferentially
forms OH. From Fig.~\ref{radicals} we see that OH is not sensitive
to H$_2$O abundance. The abundance of H$_2$O depends on the
comparative richness of cosmic C and O. Under high stellar UV
irradiation, a fraction of CO will be photodissociated. The
resulting O will react with H$_2$ to form OH, which eventually
forms H$_2$O by reacting with H$_2$. Therefore, the abundance of
OH radicals is determined by the amount of O in the system. It
does not matter whether O is in the CO or H$_2$O reservoir.

\subsection{CO$_2$ and CH$_4$ \label{CO2_CH4}}
CO$_2$ is formed via the reaction of CO and OH,
$$ \mbox{OH} + \mbox{CO} \rightarrow \mbox{CO$_2$} + \mbox{H} $$
Fig.~\ref{mr} shows the vertical profiles of CO$_2$ for our three
models. The CO$_2$ mixing ratio is enhanced in the upper
atmosphere. At pressures of $\sim 10$~nbar for Model A, the CO$_2$
mixing ratio is only about 2 orders of magnitude less than its
progenitor, CO. The CO$_2$ abundance in the model is rather
insensitive to the abundance of H$_2$O. An order of magnitude
decrease in H$_2$O results in only a factor of $\sim 3$ decrease
in CO$_2$ abundance (cf. Models A and B in Fig.~\ref{mr}).
However, CO$_2$ abundance varies approximately linearly with the
abundance of CO (cf. Models A and C in Fig.~\ref{mr}).

The formation of CH$_4$ is initiated by the downward flux of C
atoms produced in the photolysis of CO in the upper atmosphere.
This obtains the following sequence of reactions$:$
$$\mbox{CO} + \mbox{h$\nu$} \rightarrow \mbox{C} + \mbox{O},$$
$$\mbox{C} + \mbox{H$_2$} + \mbox{M} \rightarrow \mbox{$^3$CH$_2$} + \mbox{M},$$
$$2~ \mbox{$^3$CH$_2$} \rightarrow \mbox{C$_2$H$_2$} + 2\mbox{H},$$
$$\mbox{C$_2$H$_2$} + \mbox{H} + \mbox{M} \rightarrow \mbox{C$_2$H$_3$} + \mbox{M},$$
$$\mbox{C$_2$H$_3$} + \mbox{H$_2$} \rightarrow \mbox{C$_2$H$_4$} + \mbox{H},$$
$$\mbox{C$_2$H$_4$} + \mbox{H} + \mbox{M} \rightarrow \mbox{C$_2$H$_5$} + \mbox{M},$$
$$\mbox{C$_2$H$_5$} + \mbox{H} \rightarrow 2\mbox{CH$_3$},$$
$$\mbox{CH$_3$} + \mbox{H} + \mbox{M} \rightarrow \mbox{CH$_4$} + \mbox{M}.$$
Fig.~\ref{mr} shows the vertical profiles of CH$_4$ in our models.
We see that the CH$_4$ mixing ratio is increasing by a factor of
5-100 from the bottom to the 0.1~mbar level (cf. Fig.~\ref{mr}).
Above this level, CH$_4$ rapidly decreases due to
photodissociation. Some of the C is eventually converted to
CO$_2$, whose mixing ratio increases while that of CH$_4$
decreases.

The CH$_4$ mixing ratio is increased by a factor $\sim 2$ when we
lower the H$_2$O abundance by an order of magnitude (cf. Models A
and B in Fig.~\ref{mr}). We suggest this increase is due to less
UV shielding by water above. The CH$_4$ mixing ratio is decreased
by an order of magnitude when we lower the CO abundance by an
order of magnitude. The reason is that CO photolysis is the source
of the C in CH$_4$.

\subsection{H and H$_2$O \label{H_H2O}}
\begin{figure*}
\epsscale{0.7} \plotone{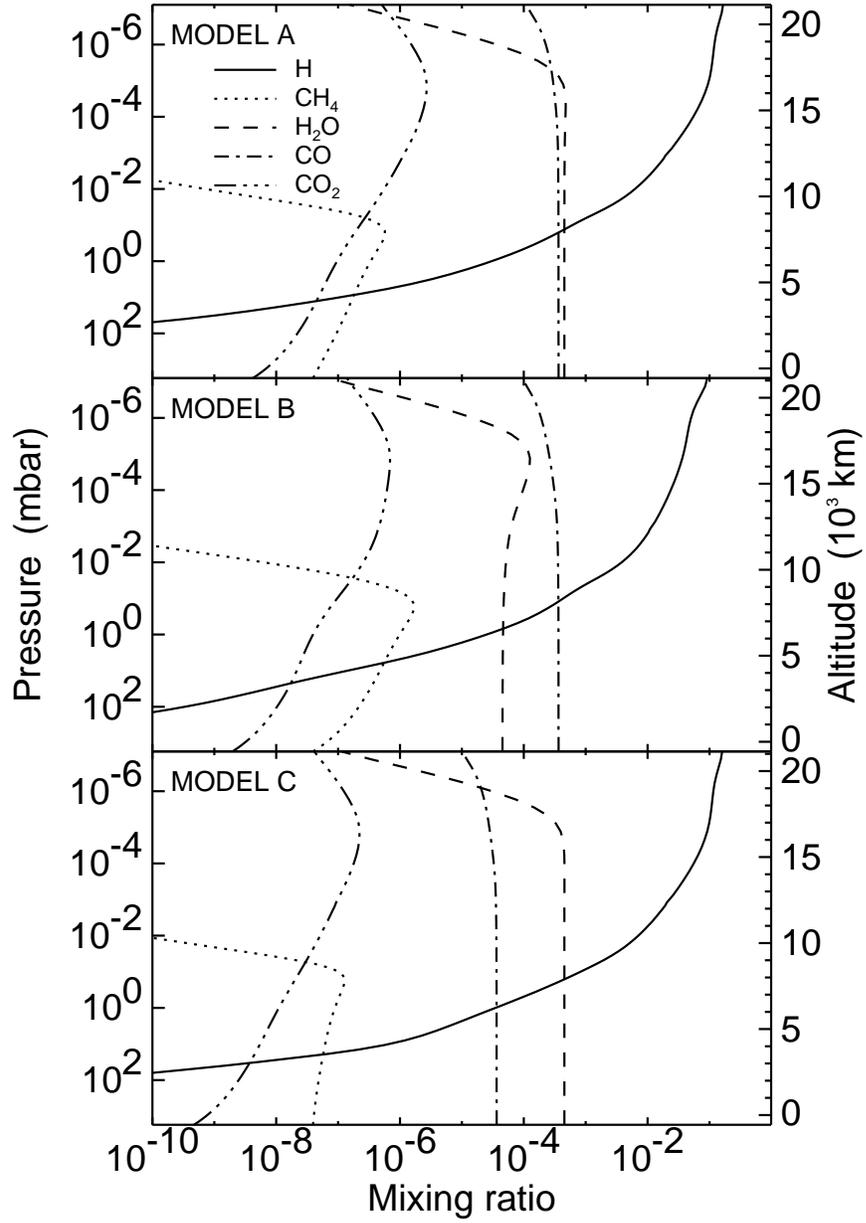} \caption[resulting mixing
ratios]{Comparison of mixing ratios of H, CH$_4$, H$_2$O, CO, and
CO$_2$ in Models A, B, and C. \label{mr}}
\end{figure*}

Fig.~\ref{mr} shows the mixing ratios for H and H$_2$O. The most
striking features are the production of H (all three models) and
the production of H$_2$O (Model B). In our 1-D model, the
production rate of atomic hydrogen is not sensitive to the exact
abundances of CO and H$_2$O. With an order of magnitude change in
the abundance of either CO or H$_2$O, the atomic hydrogen changes
by only a small factor, $\sim 1-2$. This implies the production of
H in three models is limited by the availability of UV photons.
The H production is also not sensitive to the abundance of CH$_4$.
CH$_4$ abundance has been increased to be as high as CO, and the H
mixing ratio is only changed by a small factor. A more
comprehensive discussion will be given in a separate paper (Liang
et al. 2003, in preparation). The mixing ratio of H exceeds 10\%
at the top of atmosphere. At the top of the atmosphere ($<
1$~nbar), H$_2$ will be photolyzed and is a source of H. This
atomic hydrogen will fuel the hydrodynamic loss process, as
observed by \citet{VM03} and is discussed in more detailed in the
companion paper by Parkinson et al. (2003, in preparation).

In a water-poor atmosphere (e.g., Model B), CO will be driving the
photochemical reactions to form H$_2$O$:$
$$ \mbox{CO} + \mbox{h$\nu$} \rightarrow \mbox{C} + \mbox{O}, $$
$$ \mbox{O} + \mbox{H$_2$} \rightarrow \mbox{OH} + \mbox{H}, $$
$$ \mbox{OH} + \mbox{H$_2$} \rightarrow \mbox{H$_2$O} + \mbox{H}.$$
H$_2$O and H are the net products. The produced H$_2$O will be
recycled and is an important source of OH radicals and H atoms.
Near the top of atmosphere, a large fraction of H$_2$O is
destroyed due to the high UV bombardment. However, CO is more
stable.

\begin{figure*}
\epsscale{1} \plotone{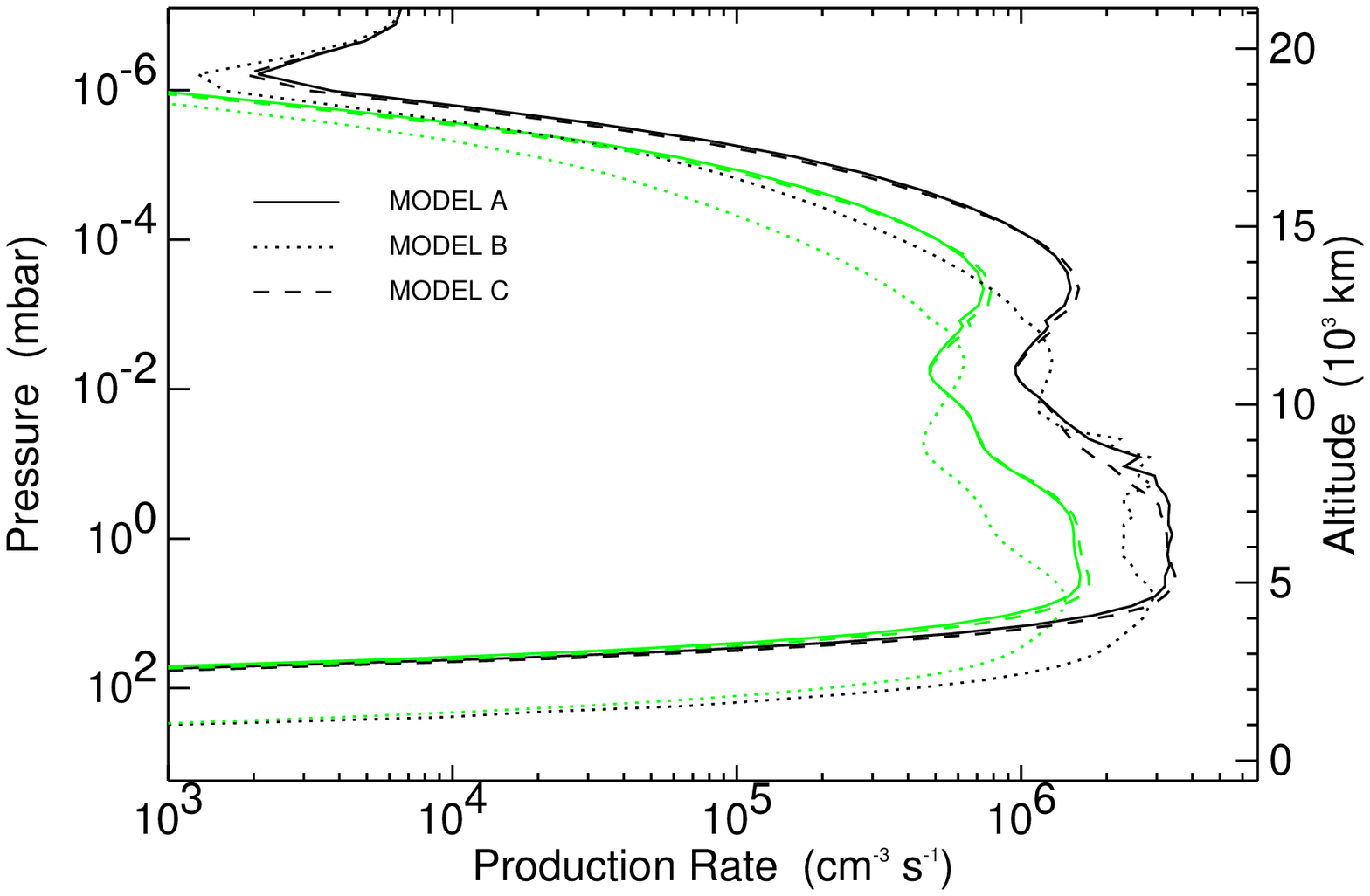} \caption[H Production Rate and
H$_2$O photolysis rate]{Production rate of H (dark lines) and
photolysis rate of H$_2$O (gray lines) in Models A, B, and C.
\label{pr}}
\end{figure*}

Fig.~\ref{pr} shows the production rate of H and the photolysis
rate of H$_2$O. The rates are not sensitive to the abundance of
CO, but are sensitive to the abundance of H$_2$O. By comparing the
H$_2$O photolysis rate with H production rate, it is evident that
production of H is mainly driven by H$_2$O photolysis and the
reaction of OH with H$_2$. Below $\sim 1$~mbar, HCO plays a role
in the removal of H via
$$ \mbox{CO} + \mbox{H} + \mbox{M} \rightarrow \mbox{HCO} + \mbox{M}, $$
$$ \mbox{HCO} + \mbox{H} \rightarrow \mbox{CO} + \mbox{H$_2$}.$$
In the upper atmosphere, H atom recombination and reactions with
CH and $^3$CH$_2$ will drive the loss of H via
$$ 2\mbox{H} + \mbox{M} \rightarrow \mbox{H$_2$} + \mbox{M},$$
and
$$ \mbox{H} + \mbox{CH} \rightarrow \mbox{C} + \mbox{H$_2$},$$
$$ \mbox{H} + \mbox{$^3$CH$_2$} \rightarrow \mbox{CH} + \mbox{H$_2$},$$
$$ \mbox{C} + \mbox{H$_2$} + \mbox{M} \rightarrow \mbox{$^3$CH$_2$} + \mbox{M}.$$

\section{CONCLUSION}

We have considered a series of possible chemical reactions using
various models for a "hot jupiter". We have shown the mechanism
for producing the atomic hydrogen. The production of H is not
sensitive to the abundances of CO, H$_2$O, and CH$_4$. Lowering
H$_2$O or CO an order of magnitude changes the concentration of H
by only a factor of $\lesssim 2$. However, the production rate of
H is sensitive to the temperature profile. A 30\% change in the
temperature will result in $\sim 50\%$ change in the H
concentration.

Our calculations show that the H mixing ratio at $\sim 1$~mbar is
$\sim 10^{-3}$ and exceeds 10\% in the top of the atmosphere.
Being less gravitationally bound, the atomic hydrogen formed at
the top of atmosphere can escape hydrodynamically as putatively
suggested by observations of \citet{VM03}. Since these close-in
gas-rich giant planets are probably tidally locked, circulation
may be important in transporting heat because the temperature
gradient can be as high as 1000~K over the entire planet, which in
turn implies wind speeds of a few km~s$^{-1}$
\citep{SG02,Cetal03}. Therefore, it is of interest to simulate the
differences in chemical processes between the day and night sides
and the global transport of heat and mass as well.

\acknowledgements We thank Dr. R. L. Shia and Dr. M. Gerstell for
helpful comments and discussions. The support of NASA Grant
NAG5-6263 to the California Institute of Technology is gratefully
acknowledged. This work represents one phase of research carried
out at the Jet Propulsion Laboratory, California Institute of
Technology, under contract to the National Aeronautics and Space
Administrations.

\end{document}